\documentstyle[12pt]{article}
\begin{document}
\def\om{\omega}
\def\omt{\tilde{\omega}}
\def\ti{\tilde}
\def\o{\Omega}
\def\bchi{\bar\chi^i}
\def\In{{\rm Int}}
\def\ba{\bar a}
\def\w{\wedge}
\def\ep{\epsilon}
\def\k{\kappa}
\def\Tr{{\rm Tr}}
\def\ST{{\rm STr}}
\def\ss{\subset}
\def\ot{\otimes}
\def\bc{{\bf C}}
\def\br{{\bf R}}
\def\de{\delta}
\def\tr{\triangleleft}
\def\al{\alpha}
\def\la{\langle}
\def\ra{\rangle}
\def\G{\Gamma}
\def\th{\theta}
\def\lm{\lambda}
\def\spj{osp(2,1)}
\def\spd{osp(2,2)}
\def\sl{sl(2,2)}
\def\jp{{1\over 2}}
\def\js{{1\over 4}}
\def\d{\partial}
\def\dz{\partial_z}
\def\dw{\partial_w}
\def\dbz{\partial_{\bar z}}
\def\dbw{\partial_{\bar w}}
\def\dbj{\partial_{b^1}}
\def\dbbj{\partial_{\bar b^1}}
\def\db2{\partial_{b^2}}
\def\dbb2{\partial_{\bar b^2}}
\def\be{\begin{equation}}
\def\ee{\end{equation}}
\def\bea{\begin{eqnarray}}
\def\eea{\end{eqnarray}}
\def\A{{\cal A}}
\def\Ai{{\cal A}_{\infty}}
\def\B{{\cal B}}
\def\T{{\cal T}}
\def\C{{\cal C}}
\def\bT{\bar{\cal T}}
\def\Z{{\cal Z}}
\def\n{{1\over n}}
\def\si{\sigma}
\def\*{\ddagger}
\def\j{\dagger}
\def\bz{\bar{z}}
\def\bw{\bar{w}}
\def\e{\varepsilon}
\def\b{\beta}
\def\ga{\gamma}
\def\bb{\bar b}
\def\bD{\bar D}
\def\bQ{\bar Q}
\def\De{ (\epsilon V+\kappa D)}
\def\bk{\bar k}
\def\ic{c^\*}
\def\cc{c^{\circ}}
\def\cic{c^{\*\circ}}
\def\IC{C^\*}
\def\CC{C^{\circ}}
\def\CIC{C^{\*\circ}}
\def\jc{{1\over c}}
\begin{titlepage}
\begin{flushright}
{}~
IML 99-03\\
hep-th/9903202
\end{flushright}

\vspace{3cm}
\begin{center}
{\Large \bf An extended fuzzy supersphere and twisted chiral
 superfields}\\
[50pt]{\small
{\bf C. Klim\v{c}\'{\i}k}\\ Institute de math\'ematiques de Luminy,
 \\163, Avenue de Luminy, 13288 Marseille, France}

\vspace{1cm}
\begin{abstract}
A noncommutative associative algebra of $N=2$ fuzzy supersphere is
introduced.
It turns out to possess a nontrivial automorphism which relates twisted
chiral to twisted anti-chiral superfields and hence makes possible
to construct  noncommutative  nonlinear $\si$-models  with extended
supersymmetry.

\end{abstract}
\end{center}
\end{titlepage}
\newpage

\section{Introduction}
Supersymmetric nonlinear $\si$-models with $N=2$ supersymmetry  in two
dimensions
are  important objects in  modern mathematical physics. They possess a
very rich
structure interesting by itself and  find also applications, for instance,
 in superstring theory.
 It is
a well-known fact that the models with $N=1$ supersymmetry can be
constructed
for an arbitrary geometry of the target space. However, the $N=2$ case
 requires
the target space to be K\"ahler \cite{Bel} if we consider the case without
 torsion.

There exists a very convenient description of the $N=2$ $\si$-models based
 on the $N=2$
superspace. In this paper, we shall show that the $N=2$ superspace can be
 constructed
also on the noncommutative sphere. More precisely, we shall construct a
 noncommutative
$N=2$ supersphere. Note that the notation $N=1$ or $N=2$ refers usually to
the Poincare-like
superalgebras in which the anticommutators of the supercharges are the
 generators of translations
of the underlying bosonic space. We shall see soon, however, that due to
 the fact that the
two-sphere is conformally flat we can keep this terminology  also for
 spherical worldsheets.

Noncommutative geometry \cite{Con} is the generalization of the
ordinary geometry in which an algebra of functions which encodes the
geometry
of an ordinary space is replaced by certain  noncommutative algebra.
As an example we take  a noncommutative (or fuzzy) sphere which is an
object introduced by several
researchers in the past \cite{Ber,Hop, Mad, GKP2} with various
motivations. Berezin himself
has quantized the standard round symplectic structure on the two-sphere
and he found that
this can be done only for integer values of the inverse Planck constant.
 For example if
$h=1/n$, the quantized algebra of observables (=the fuzzy sphere)
coincides simply with the algebra of $(n+1)\times
(n+1)$ matrices. When $n\to\infty$ the size of the algebra approaches
infinity; in fact, one
recovers the standard algebra of functions on the commutative sphere.
Effectively, the quantization cuts off the  large angular momenta. This
 fact lead
independently several authors \cite{Hop,Mad,GKP2} to use the fuzzy
sphere
as a regularization of fields theories formulated on the ordinary sphere.

It turns out that this
regularization has an important advantage of preserving the standard
$SO(3)$ invariance of the
ordinary sphere. This is a quite remarkable fact because the regulated
theory contains only
 a finite number of degrees of freedom and, even more importantly, the
regulated
sphere continues to be a geometric object so it makes sense to formulate
theories
 non-perturbatively directly on it.

The list of the virtues of the fuzzy regularization is not exhausted by
 the $SO(3)$ invariance
and the finite number of degrees of freedom. In fact, one can introduce
fuzzy monopole configurations
\cite{GKP3,Bal} and, perhaps even more remarkable, to regulate
supersymmetric \cite{GKP2}
and supersymmetric gauge theories \cite{K2} while manifestly
preserving supersymmetry, supergauge symmetry
 and the finite number of degrees of freedom. It is indeed the
purpose of this paper to show that  models with  {\it extended}
 supersymmetry are also regularisable
by the method.

In section 2, we introduce the extended $N=2$ supersphere and its
noncommutative deformation.
Moreover, we shall identify a nontrivial automorphism of the structure
 which will prove
very useful in constructing $N=2$ theories.
Section 3 presents the construction of the commutative and noncommutative
$N=2$ supersymmetric
nonlinear $\sigma$-models on the sphere.

\section{$N=2$ fuzzy supersphere}

An $N=1$ fuzzy supersphere has been constructed in \cite{GKP2} with a goal
 to regularize
 $N=1$ supersymmetric nonlinear $\si$-models.
The reader may find an alternative more concise description of the
structure in \cite{K2}.
 In the $N=2$ case, the construction begins in a similar way than in the
 $N=1$ one but there is a point
of depart in which new (and welcome) structural ingredients enter. Here
 are the details.

Consider the algebra of polynomial functions on the complex $C^{2,2}$
 superplane, i.e.
 algebra
generated by finite sums of monomials in bosonic variables
$\bar\chi^{\al},\chi^{\al}, \al=1,2$ and in
fermionic ones $\bar a^{\al},a^{\al}, \al=1,2$.
The algebra is equipped with the super-Poisson bracket
\be \{f,g\}=
\d_{\chi^{\al}}f\d_{\bar\chi^{\al}}g-\d_{\bar\chi^{\al}}f\d_{\chi^{\al}}g+
(-1)^{f+1}[\d_{a^{\al}} f\d_{\bar a^{\al}}g+
\d_{\bar a^{\al}} f\d_{a^{\al}} g].\ee
and  with the graded involution \cite{SNR}
\be (\chi^{\al})^{\*}=
\bar\chi^{\al},~(\bar\chi^{\al})^{\*}=\chi^{\al},
~(a^1)^{\*}=\bar a^1,~ (a^2)^\*=-\bar a^2,~ (\bar a^1)^\*=-a^1,
~ (\bar a^2)^{\*}=a^2,\ee
satisfying  the following properties
\be (AB)^\*=(-1)^{AB}B^\* A^\*,\quad  (A^\*)^\*=(-1)^A A.\ee
 We can now apply the (super)symplectic reduction with respect
to a moment map $\bar\chi^{\al}\chi^{\al} +\bar a^{\al} a^{\al} -1$.
 The result is a
 smaller algebra
$\A_{\infty}$, that by definition consists of all functions $f$ with the
property
\be \{f,\bar\chi^i\chi^i+\bar a^{\al} a^{\al}-1\}=0.\ee
Moreover, two functions obeying (4)
are considered to be equivalent if they differ just by a product
of $(\bar\chi^{\al}\chi^{\al}+\bar a^{\al}  a^{\al}-1)$ with some other
such function.
The smaller algebra $\A_{\infty}$ (the reason for using of the subscript
$\infty$ will become clear soon)
 will be referred to as the algebra of superfunctions on an $N=2$
supersphere\footnote{Note
that in case we did not consider the fermionic variables
$\bar a^{\al},a^{\al}$, we would obtain,
as the result of the symplectic reduction,
the algebra of functions on the standard bosonic sphere.
 In case of considering only one pair of fermionic variables $\bar a,a$
we would
obtain the $N=1$ supersphere.}. It will
be sometimes more convenient to work with a different
 parametrization  of $\A_{\infty}$,
using  the following coordinates
\be z={\chi^1\over\chi^2},
\quad \bar z ={\bar\chi^1\over\bar\chi^2},
\quad b^{\al}={a^{\al}\over \chi^2},\quad \bar b^{\al}={\bar a^{\al}\over
\bar \chi^2}.\ee

 The Poisson bracket (1) then becomes
$$ \{f,g\}=(1+\bz z + \bb^{\al} b^{\al})[(1+\bz z)(\dz f \dbz g - \dbz f
\dz g)$$
$$ \bb^{\b} z
((-1)^f\dz f \d_{\bb^{\b}} g - \d_{\bb^{\b}} f \dz g) + b^{\b}\bz
(\d_{b^{\b}} f \dbz g -
(-1)^f\dbz f \d_{b^\b} g)$$
\be +(-1)^{(f+1)}(-\bb^{\b}b^\ga+\de^{\b\ga})) (\d_{b^\ga}
f \d_{\bb^\b}g +
\d_{\bb^\b}f \d_{b^\ga} g)]. \ee
 A natural Berezin integral on $\A_{\infty}$ can be written as
\be I(f)=~{1\over (2\pi i)^2}
\int d\bar\chi^1\w d\chi^1\w d\bar\chi^2\w d\chi^2 \w
d\bar a^1 \w da^1 \w d\bar a^2\w da^2~\delta(\bar\chi^\al \chi^\al +
\bar a^\al a^\al-1) f.\ee
It can be rewritten as
\be I(f)\equiv
{1\over 2\pi i}\int d\bz \w dz \w d\bb^1 \w db^1 \w d\bb^2\w db^2 f.\ee
(Note $I(1)=0$.)

Strictly speaking, the generators $\bz,z,\bb^\al,b^\al$ are not elements
of the algebra $\A_{\infty}$.
What is true is that $\Ai$ is (finitely) linearly generated by the
functions
of the  following form
\be  {\bz^{\bar k} z^k (\bb^1)^{\bar l^1} (b^1)^{l^1} (\bb^2)^{\bar l^2}
(b^2)^{l^2}\over
(1+\bz z +\bb^\al b^\al)^m},
\quad \bar k +\bar l^1+\bar l^2, k+l^1+l^2 \leq m,\ee
where $ k,\bar k,l^\al,\bar l^\al,m$ are non-negative integers.
It is not difficult to understand the form (9) of the elements of $\Ai$.
 Indeed,
we first note
that $\bz,z,\bb^\al,b^\al $ can be also interpreted as a local chart
 coordinates of the
 $N=2$-supersphere
obtained by the stereographic projection from the north pole. If we do the
 projection from the
south pole, we obtain a complementary chart with local coordinates
$\bar w,w,\bar b_w^\al,b_w^\al$.
A transition rule on the overlap of the two charts reads
\be w=1/z,\quad \bw =1/\bz,\quad b_w^\al=b^\al/z,\quad
\bar b_w^\al=\bb^\al/\bz.\ee
It is now a simple matter to check that the functions of the form (9)
 will transform into
\be  {\bar w^{m-\bar k-\bar l^1-\bar l^2} w^{m-k-l^1-l^2}
 (\bb_w^1)^{\bar l^1} (b_w^1)^{l^1}
(\bb_w^2)^{\bar l^2} (b_w^2)^{l^2}\over
(1+\bar w w +\bb_w^\al b_w^\al)^m}.\ee
Since $0\leq m-\bar k-\bar l^1-\bar l^2\leq m$ and
 $0\leq m- k- l^1- l^2\leq m$ for
 $\bar k +\bar l^1+\bar l^2, k+l^1+l^2 \leq m$;
$k,\bar k,l^\al,\bar l^\al,m\geq 0$
 we see that the elements of $\Ai$ are form-invariant with respect
to the coordinate transformation (10).

The reason to use the coordinates $\bz,z,\bb^\al,b^\al$ is
 simple: they  will enable
us to establish a connection between standard $N=2$ supersymmetric
 nonlinear $\si$-models
defined on the flat Euclidean space and their counterparts on the $N=2$
supersphere. In fact, we shall
see that the flat models in the coordinates $\bz,z,\bb^\al,b^\al$ and the
spherical models in the
same coordinates have the same field theoretical action! They $differ$,
however, in the sense that
the algebras of the superfields in both cases are different. In the flat
case the superfield is an element
of the algebra of superfunctions on the Euclidean $N=2$ superspace while
 in the spherical case
the superfield is an element of $\Ai$.

Let us now introduce a Lie superalgebra $\T$ which will
turn out to contain all relevant structure of the $N=2$ nonlinear
$\sigma$-models on the sphere.
It has seven
 even generators $R_{\pm},R_3,Z_{\pm},Z_3,C$
and eight odd ones $C_{\pm},C_{\pm}^\*,C_{\pm}^{\circ},
C_{\pm}^{\*\circ}$.
We denote the corresponding Hamiltonians by small characters; they are
 given by
\be  r_3=\jp(\bar\chi^1\chi^1-\bar\chi^2\chi^2),\quad r_+=\bar\chi^1\chi^2,
\quad  r_-=\bar\chi^2\chi^1;\ee
\be z_3=\jp(\bar a_1 a_1-\bar a_2 a_2),\quad z_+=\bar a_1 a_2,
\quad z_-=\bar a_2 a_1;\ee
\be c=\bar\chi^1\chi^1 +\bar\chi^2\chi^2 +\bar a^1 a^1+\bar a^2 a^2;\ee
\be c_+=\bar a^2\chi^1 +\bar\chi^2 a^1,\quad c_-=-\bar a^2\chi^2+
\bar\chi^1 a^1;\ee
\be \ic_+=\bar a^1\chi^2 +\bar\chi^1 a^2,\quad \ic_-=\bar a^1\chi^1-
\bar\chi^2 a^2;\ee
\be  \cc_+=\bar a^1\chi^1 +\bar\chi^2 a^2,\quad \cc_-=-\bar a^1\chi^2+
\bar\chi^1 a^2;\ee
\be \cic_+=\bar a^2\chi^2 +\bar\chi^1 a^1,\quad \cic_-=\bar a^2\chi^1-
\bar\chi^2 a^1.\ee
We should remember that these Hamiltonians are actually preimages of the
true
Hamiltonians in the process of the symplectic reduction. Since they
 anyway commute with the moment map
$(c-1)$ it is possible and technically preferable to work with them.
 A reader who wishes to work
directly with expressions in terms of $\bz,z,\bb^\al,b^\al$ coordinates
 can simply use the equations
(12) -(18) and the following relation
\be {1\over \bar\chi^2\chi^2}=1+\bz z+\bb^1b^1+\bb^2 b^2.\ee
One obtains, for example,
\be c_+={z\bb^2+b^1\over 1+\bz z+\bb^1b^1+\bb^2 b^2},\quad
 c_-={\bz b^1 -\bb^2\over 1+\bz z+\bb^1b^1+\bb^2 b^2}\ee
and so on for all Hamiltonians (12) -(18). In particular, the Hamiltonian
$c$
 becomes simply
\be c=1.\ee The last equation does not mean, however, that $C$ gets
 detached from the superalgebra
$\T$. It rather means that $\T$ is the  central extension of $\T/C$ by $C$.

The (graded) commutation relations of the superalgebra $\T$ are given by
the Poisson brackets
of the Hamiltonians (12)-(18). Though this is a correct definition,
 we prefer
to give an explicit  list
of nonvanishing commutators  because many results of this paper depend
directly on them. Here they are
\be [R_3,R_{\pm}]=\pm R_{\pm},\quad   [Z_3,Z_{\pm}]=\pm Z_{\pm},\quad
[R_+,R_-]=2R_3,\quad
 [Z_+,Z_-]=2Z_3;\ee
\be [R_3,C_{\pm}]=\mp\jp  C_{\pm},~[R_3,\CC_{\pm}]=\mp\jp\CC_{\pm},~
[R_3,\IC_{\pm}]=\pm\jp  \IC_{\pm},
~[R_3,\CIC_{\pm}]=\pm\jp  \CIC_{\pm};\ee
\be  [R_{\pm},C_{\pm}]=  C_{\mp},~[R_{\pm},\CC_{\pm}]=\CC_{\mp},~
[R_{\pm},\IC_{\mp}]=-\IC_{\pm},
~[R_{\pm},\CIC_{\mp}]=- \CIC_{\pm};\ee
\be [Z_3,C_{\pm}]=-\jp  C_{\pm},~[Z_3,\CC_{\pm}]=\jp\CC_{\pm},~
[Z_3,\IC_{\pm}]=\jp  \IC_{\pm},
~[Z_3,\CIC_{\pm}]=-\jp \CIC_{\pm};\ee
\be [Z_+,C_{\pm}]=\pm\IC_{\mp},~[Z_+,\CIC_{\pm}]=\mp\CC_{\mp},~
[Z_-,\CC_{\pm}]=\pm  \CIC_{\mp},
~[Z_-,\IC_{\pm}]=\mp C_{\mp};\ee
\be [C_{\pm},\CC_{\pm}]_+=\pm 2R_{\mp},~[\IC_{\pm},\CIC_{\pm}]_+=
\pm 2R_{\pm},
~[C_{\pm},\CIC_{\pm}]_+= 2Z_-,~[\CC_{\pm},\IC_{\pm}]_+=2Z_+;\ee
\be [C_{\pm},\CC_{\mp}]_+= [\IC_{\pm},\CIC_{\mp}]_+=2(R_3\mp Z_3),~
[C_{\pm},\IC_{\pm}]_+= [\CC_{\pm},\CIC_{\pm}]_+=C.\ee
We note that the commutation relations of $\T/C$ coincide with those of
the anomaly free
subalgebra of the $N=4$ super-Virasoro algebra \cite{Fr}.

Let us define an automorphism $\circ$ of the algebra $\T$ which plays  a
crucial role in our
construction. It is easy to verify that the commutation relations of $\T$
are invariant if
\be R_{\pm}^\circ=R_{\pm},\quad R_3^\circ=R_3,\quad  Z_+^\circ =Z_-,\quad
 Z_3^\circ=-Z_3,
\quad C^\circ=C.\ee
The action of the automorphism on the odd generators is given by the
 notation itself and
by the claim that the automorphism is involutive i.e. it squares to the
 identity map.

It is a matter of a simple inspection to see that the associative algebra
 $\Ai$, which defines
the $N=2$ supersphere, is linearly and multiplicatively
generated by four odd variables $c_{\pm},\ic_{\pm}$ and three even ones
 $l_{\pm},l_3$,
defined by\footnote{Although $c=1$ in $\Ai$, it is useful to indicate $c$
in (30) and (31) because
these formulae hold also in the noncommutative case where $c\neq 1$.}
\be l_3=r_3-{1\over 2c}\ic_+ c_++{1\over 2c}\ic_- c_-;\ee
\be l_{\pm}=r_{\pm}+\jc\ic_{\pm}c_{\mp}.\ee
Note that
\be l_+^\*=l_-,\quad l_3^\*=l_3\ee
and that $l_{\pm},l_3$ are not independent variables as they are subject
to the following
relation
\be l_3^2+l_+l_-=1/4.\ee
But the relations (32) and (33)  characterize the ordinary bosonic sphere!
 Moreover, the
only nonvanishing Poisson brackets among the generators
 $l_{\pm},l_3,c_{\pm},\ic_{\pm},c$
turn out to be
\be \{l_3,l_{\pm}\}=\pm l_{\pm},\quad \{l_+,l_-\}=2l_3;\ee
\be \{c_{\pm},\ic_{\pm}\}=c.\ee
In other words, $l$'s and $c$'s completely decouple and we see that the
algebra $\Ai$
is  a direct product of the algebra $\B_{\infty}$ of the functions on the
 ordinary sphere
and of the Grassmann algebra $Gr_4$ with four generators $c_\pm,\ic_\pm$.
This direct product
 concerns not only the associative multiplication but also the Poisson
structure.
The  immediate conclusion of those facts is that it is very easy to
quantize the $N=2$
supersphere.  The corresponding noncommutative
algebra $A_n$ is simply the ordinary bosonic fuzzy sphere
\cite{Hop,Mad,GKP2} tensored
with a Clifford algebra $Cf_4$ with four generators  $c_{\pm},\ic_{\pm}$.
 Here and
in what follows we shall often use the same symbol for non-deformed
 generators and for
their deformed counterparts . It should be clear from the context which
usage we have in mind.

We remind that the bosonic fuzzy sphere is a $(n+1)\times(n+1)$
dimensional matrix algebra where the integer parameter $n$ plays role of the
inverse Planck constant \cite{GKP2,K,K2}.   Since the Poisson brackets
(34) and (35) are to be replaced
by commutators scaled by the inverse Planck constant, we get the
following  commutation relations
for the noncommutative generators  $l_{\pm},l_3,c_{\pm},\ic_{\pm}$ and
 $c$ of the  $N=2$
fuzzy supersphere
\be [l_3,l_{\pm}]=\pm{1\over n} l_{\pm},\quad [l_+,l_-]={2\over n}l_3;\ee
\be [c_{\pm},\ic_{\pm}]_+=\n c.\ee
Moreover, we define the graded involution $\*$ in the noncommutative case
by (32) on $l_\pm,l_3$ and by the notation and the second property (3) on
$c_\pm,\ic_\pm$.
It is easy to find the explicit forms of the matrices
 $l_{\pm},l_3,c_{\pm},\ic_{\pm}$ and $c$:
\be l_\pm=\n( L_\pm\ot 1),\quad l_3=\n( L_3\ot 1),
\quad c=(1+\n)(1\ot 1);\ee
\be c_\pm={1\over \sqrt{n}}(1\ot \gamma_\pm),\quad \ic_\pm=
{1\over \sqrt{n}}(1\ot\gamma_\pm^\*),\ee
where the first entry of the tensor product corresponds to the
bosonic fuzzy sphere and the second entry to the Clifford algebra.
$L_\pm,L_3$ are generators of $su(2)$  Lie algebra in the representation
with spin $n/2$
and $\gamma$'s and $\gamma^\*$'s are standard Dirac matrices with respect
to the  Euclidean
metric in four dimensions normalized according to
\be [\gamma_\pm,\gamma_\pm^\*]_+=c=1+1/n.\ee
Note that due to the tensor product structure of the $N=2$ supersphere,
 the normalization
of the central term $c$
must be $1$ in the limit $n\to\infty$ but it is otherwise a free parameter
 of the construction.
 It is the choice $c=1+1/n$ which makes possible to construct $N=2$
supersymmetric $\sigma$-models
on the $N=2$ fuzzy supersphere.

Let us study the properties of the algebra $\A_n$.
It turns out that we shall need a non-commutative  analogue of the
 automorphism $\circ$.
 Actually, we defined $\circ$
as the automorphism of the Lie superalgebra $\T$ and not yet of $\Ai$.
On the other hand,
$\circ$ can be directly defined to act on the whole algebra $\Ai$ as
the morphism of the (graded) multiplication.
For instance,
\be (l_3c_+)^\circ =l_3^\circ \cc_+.\ee
Since we  know that  the Poisson bracket
 (6) is compatible
with the associative multiplication in $\Ai$, we conclude  that
 $\circ$ is the authomorphism of the both associative and Poisson
 structure of $\Ai$.

We can equally well use another set of generators for describing the
algebra $\Ai$, namely,
the set  $\cc_{\pm},\cic_{\pm}$ and  $l_{\pm}^\circ,l_3^\circ$.
Of course, $l_{\pm}^\circ,l_3^\circ$
are given by
\be l_3^\circ=r_3-{1\over 2c}\cic_+ \cc_++{1\over 2c}\cic_- \cc_-;\ee
\be l_{\pm}^\circ=r_{\pm}+\jc\cic_{\pm}\cc_{\mp}\ee
and they also turn out  to fulfil the relations
\be l_+^{\circ\*}=l_-^\circ,\quad l_3^{\circ\*}=l_3^\circ\ee
and
\be (l_3^\circ)^2+l_+^\circ l_-^\circ=1/4.\ee
Moreover, the
only nonvanishing Poisson brackets among the generators
$l_{\pm}^\circ,l_3^\circ,\cc_{\pm},\cic_{\pm}$
are as follows
\be \{l_3^\circ,l_{\pm}^\circ\}=\pm l_{\pm}^\circ,
\quad \{l_+^\circ,l_-^\circ\}=2l_3^\circ;\ee
\be \{\cc_{\pm},\cic_{\pm}\}=c.\ee
The relations (46) and (47) are actually the direct consequences
 of the fact that
$\circ$ is the automorphism
of the Poisson algebra $\Ai$. Nevertheless, we prefer to state them
explicitely in order
to stress the equal footing of the two sets of generators.

Of course, we can now construct the noncommutative  deformation of the
 $N=2$ supersphere,
by quantizing the set of the new circled generators. Thus, the $N=2$ fuzzy
supersphere will be again
nothing but the tensor product of the  bosonic fuzzy sphere with the
Clifford algebra $Cf_4$.
A question arises: How the uncircled and the circled fuzzy superspheres
fit together?
 Let us look for a key
for answering
this question  in the commutative case $\Ai$,
where the circled variables can be  written in terms of the  non-circled
ones as follows
\be \cc_{\pm}=
\jc(2l_3\ic_{\mp}\pm 2l_\mp \ic_\pm\pm\jc\ic_\mp[\ic_\pm,c_\pm]);\ee
\be \cic_{\pm}=
\jc(2l_3c_{\mp}\pm 2l_\pm c_\pm\pm\jc c_\mp[\ic_\pm,c_\pm]);\ee
\be l_3^\circ=l_3+{1\over 2c}\ic_+ c_+-{1\over 2c}\ic_- c_-
-{1\over 2c}\cic_+ \cc_++{1\over 2c}\cic_- \cc_-\ee
\be l_{\pm}^\circ=l_{\pm}-\jc\ic_{\pm}c_{\mp}+\jc\cic_{\pm}\cc_{\mp}.\ee
Now we  take the formulae (48)-(51) as the $definition$ of the circled
variables in the
 non-commutative case where
 $l_\pm,l_3,c_\pm;\ic_\pm,c$ are given by (38) and (39).

The operator formulae (48) and (49)
are remarkable since  they involve cubic terms
in the old uncircled generators. This causes that the usual  ordering
 problem leads in this case
to an $operator$ rather than a $c$-number ambiguity. Indeed, writing the
 cubic terms in (48) and (49) requires
the fixing of a certain ordering; in fact, the commutators
 (not anticommutators!)
 $[\ic_\pm,c_\pm]$ in (48) and (49) do the job. A slight change of the
 ordering in any of the definitions
 (48),(49)
would completely destroy a crucial property of this maps, namely,
the circled variables
 fulfil exactly the same properties as the noncircled ones.
Explicitely,
\be [l_3^\circ,l_{\pm}^\circ]=\pm \n l_{\pm}^\circ,
\quad [l_+^\circ,l_-^\circ]={2\over n}l_3^\circ;\ee
\be [\cc_{\pm},\cic_{\pm}]_+=\n c\ee
and all remaining graded commutators vanish.
 Remind that the relations (52) and (53)
 are not postulated but they are derived
from the relations (36) and (37) and
the definitions (48) and (49).
The normalization and reality of $l_3^\circ,l_{\pm}^\circ$ is also correct
 since
one can verify that
\be (l_3^\circ)^2+l_+^\circ l_-^\circ= l^2+l_+ l_-=1/4 +1/2n\ee
and
\be l_+^{\circ\*}=l_-^\circ,\quad l_3^{\circ\*}=l_3^\circ.\ee

Thus we conclude that the uncircled $N=2$ fuzzy supersphere is the same
 thing as the circled one.
The mapping
 $\circ$ preserve the commutation relations among the generators
therefore it can be extended to the whole supersphere as the  automorphism
 of its associative
product. Moreover, $\circ$
 is an involutive automorphism since (51) and (52) are
 manifestly involutive and a
tedious computation
shows that  the definitions (48) - (51) imply
\be c_{\pm}=
\jc(2l_3^\circ\cic_{\mp}\pm 2l_\mp ^\circ\cic_\pm\pm\jc\cic_\mp[\cic_\pm,
\cc_\pm]);\ee
\be \ic_{\pm}=\jc(2l_3^\circ\cc_{\mp}\pm 2l_\pm^\circ \cc_\pm\pm\jc
\cc_\mp[\cic_\pm,\cc_\pm]).\ee

For a completeness, we give explicit formulae for the even Hamiltonians
$r_\pm,r_3,z_\pm,z_3$
in terms of the generators $l_3,l_\pm,c_\pm,\ic_\pm$ and
$l_3^\circ,l_\pm^\circ,\cc_\pm,\cic_\pm$.
They are valid in both commutative ($n\to\infty$) and noncommutative
(finite $n$) cases:
\be r_3=l_3+{1\over 2c}\ic_+ c_+-{1\over 2c}\ic_- c_-=l_3^\circ
+{1\over 2c}\cic_+ \cc_+-{1\over 2c}\cic_- \cc_-;\ee
\be r_\pm =l_{\pm}-\jc\ic_{\pm}c_{\mp}=
l_\pm^\circ -\jc\cic_{\pm}\cc_{\mp};\ee
\be  z_3={1\over 2c}\ic_+ c_++{1\over 2c}\ic_- c_- -{1\over 2n}=
{1\over 2n}
-{1\over 2c}\cic_+ \cc_+-{1\over 2c}\cic_- \cc_- ;\ee
\be z_+=\jc\ic_-\ic_+=\jc\cc_+\cc_-,\quad z_-=
\jc c_+c_-=\jc \cic_-\cic_+.\ee

The construction of the involutive automorphism $\circ$
is the main result of this section. In what follows, we shall  always
enjoy  a  freedom of choosing to  work
in one of the two $\circ$ related equivalent parametrization of the
 $N=2$ fuzzy supersphere.

\section{$N=2$ nonlinear $\sigma$-models}

\subsection{The commutative case}

The basic fact of life in the $N=2$ flat Euclidean superspace is that a
 Lagrangian
density of a field theoretic model does not involve derivatives. All
dynamics is encoded
in constraints imposed on $N=2$ superfields in a way compatible with the
$N=2$ supersymmetry.
For example, the Lagrangian of  an  $N=2$ supersymmetric $\si$-model on
 the Euclidean
plane is given by
\be S=\int d\bz dz d\bb^1 db^1d\bb^2 db^2 K(\bar\Phi \Phi).\ee
Here $\Phi(\bz,z,\bb^1,b^1,\bb^2,b^2)$ and
$\bar\Phi(\bz,z,\bb^1,b^1,\bb^2,b^2)$ are superfields
on the plane. They are subject to the
 following constraints
\be D_+\Phi=\bar D_-\Phi=0; \quad D_-\bar\Phi=\bar D_+\bar\Phi=0\ee
and $K(\bar\Phi\Phi)$ is the  K\"ahler potential of a target  K\"ahler
 manifold   with
complex coordinates $\bar\Phi,\Phi$. The supersymmetric covariant
derivatives are defined as
\be D_+=\db2 +b^1\dz,\quad D_-=\dbj +b^2\dz,\quad \bD_+=\dbb2+\bb^1\dbz,
 \quad \bD_-=\dbbj+\bb^2\dbz.\ee

Note that the flat measure in the integral (62) coincides with the $N=2$
"round" measure (8).
Because of this fact, the model  (62) can  be reinterpreted as a model
on the $N=2$ supersphere. For this interpretation, it is sufficient to
declare that both $\bar\Phi$
and $\Phi$ are not the superfields on the plane
but but they are rather elements of the algebra $\Ai$ i.e. of the algebra
 of the superfunctions on
the $N=2$
supersphere. More precisely, the superfields are linear combinations of
the elements of $\Ai$
of the form (9) with coefficients being ordinary numbers, when
 $\bar l^1+\bar l^2 +l^1 +l^2$ is
even and Grassmann numbers when $\bar l^1+\bar l^2 +l^1 +l^2$ is odd.
 These Grassmann numbers
anticommute with the odd generators of $\Ai$. As a result, the
superfields $\bar\Phi,\Phi$ are even. This remark is important when we
 calculate
the Poisson brackets involving the superfields or when we use the graded
involution.

The constraints (63) turn out to be equivalent to
\be \{c_\pm,\Phi\}=0, \quad \{\cc_\pm,\bar\Phi\}=0,\ee
where $\{.,.\}$ is the "round" Poisson bracket (6) and the Hamiltonians
$c_\pm,\cc_\pm$ are given
in (15) and (17).
The constraints (63) or (65) define so-called twisted chiral and
twisted
anti-chiral fields, respectively. In order to have another viable set of
 Poisson bracket
constraints, giving so-called  chiral and anti-chiral superfields
\cite{Bel}, we would have to change
the symplectic structure (6). This is easy but we shall not discuss it in
 this paper because
the resulting picture is completely analoguous to the twisted one. We just
 remark, that from
the point of view of the  Poisson bracket (6) on the $N=2$ supersphere,
the twisted
fields are more "natural" than the untwisted ones.

 There is an  inconspicuous but, in fact, an important detail that
concerns the (graded) involution
in (62)  denoted by a bar. It acts on the
generators $\bz,z,\bb^1,b^1,\bb^2,b^2$ following the notation itself .
This involution $is$ $not$ the
 same as the one denoted by
$\*$ in section 2  (cf. (2)), although they coincide on the  bosonic
variables $\bz,z$.
In fact,  $\*$ is rather a world-sheet involution. It is with respect
to $\*$ that the generators
$l_\pm,l$ or $l_\pm^\circ,l^\circ$ fulfil the correct reality conditions
(32) of the bosonic
generators of the (fuzzy) sphere. On the other hand, the bar involution
sets the reality
properties of fermionic fields if the supersymmetric action is written in
 components. These
reality properties
propagate to the quantization of field theoretical model and define an
involution on the Hilbert
space of the quantum field theory. We remark that all this is also a
standard flat space
supersymmetric story although many authors do not provide a detailed
discussion of various
involutions in game. Their approach is simple and pragmatic, once a
 Lagrangian is worked out
in components, an
involution on fermions  is set  which makes the action real. In our case,
we have to be more careful
since an experience \cite{GKP2,K} teaches us that only superfields as
 whole are deformable; in other
words, the notion of the component fields may loose sense after the
noncommutative deformation.

Once we have defined the $\sigma$-model (62) on the commutative $N=2$
supersphere, it is natural to ask
what is its algebra of supersymmetry. There is a huge formal supersymmetry
 algebra
of the theory (62) known as the $N=2$ super-de-Witt  algebra
 (whose central extension is $N=2$
Virasoro algebra \cite{Fr}). It is actually defined as the Lie
superalgebra of vector fields
that preserve the constraints (63) \cite{Sch} and, explicitely, it is
generated by even chiral
(anti-chiral)
vector fields $L_k,J_k,k\in{\bf Z}$ ($\bar L_k,\bar J_k,k\in{\bf Z}$) and
 odd chiral (anti-chiral) ones
$G^{\pm}_{k+\jp},k\in{\bf Z}$ ($\bar G^{\pm}_{k+\jp},k\in{\bf Z}$). They
are given by
\be L_k=z^{-k+1}\dz +\jp(-k+1)z^{-k}(b^1\dbj+b^2\db2);\ee
\be J_k=z^{-k}(b^1\dbj -b^2\db2);\ee
\be G^+_{k+\jp}=(z^{-k}\db2+kz^{-k-1}b^1b^2\db2 -z^{-k}b^1\dz);\ee
\be G^-_{k+\jp}=(z^{-k}\dbj+kz^{-k-1}b^2b^1\dbj -z^{-k}b^2\dz).\ee
The barred generators are given by the same formulas with
$\bz,\bb^1,\bb^2$ replacing $z,b^1,b^2$.
It turns out that only eight chiral generators
$L_{\pm 1},L_0,J_0,G^\pm_{\pm\jp}$ and eight anti-chiral
ones $\bar L_{\pm 1}\bar L_0,\bar G^\pm_{\pm\jp}$ preserve the algebra
 $\Ai$ of the superfields
on $N=2$ supersphere. Obviously, they form a Lie subalgebra
of the full de Witt algebra. It therefore
 seems that the algebra of supersymmetry
 of the model on the sphere
has sixteen complex dimensions. However, it is not so because we have to
impose two further
condition which the supersymmetry algebra has to fulfil:
\vskip1pc
\noindent  1) Since we are interested in
the  noncommutative deformation of the $N=2$ $\sigma$-model (62)  we
have
to consider  only those
generators which act by means of the Poisson bracket (6).
This means that they are the Hamiltonian vector fields and, in the
noncommutative case,
they will act via the commutators.
This  reduces the supersymmetry algebra to an eight dimensional Lie
superalgebra $spl(2,1)$.
It is generated by four even generators $R_\pm,R_3,Z_3$ and four
odd ones $\IC_\pm,\CIC_\pm$. Explicitly,
\be \IC_+=G^-_\jp+\bar G^+_{-\jp},~ \IC_-=G^-_{-\jp}-\bar G^+_\jp,
~\CIC_+=G^+_\jp+\bar G^-_{-\jp},
~ \CIC_-=G^+_{-\jp}-\bar G^-_\jp;\ee
\be R_+=-L_1-\bar L_{-1},~ R_-=\bar L_1+L_{-1},~ R_3=\bar L_0-L_0,~ Z_3=
\jp(\bar J_0-J_0).\ee
Needless to say, the Hamiltonians of these generators of $spl(2,1)$ are
$r_\pm,r_3,z_3$ and
 $\ic_\pm,\cic_\pm$ of section 2, Eqs (12),(13),(16) and (18).
 Clearly, $spl(2,1)$
is the Lie  subalgebra of $\T$ hence its  commutation relations are
contained in (22) - (28).
\vskip1pc
\noindent 2)
We require that a supersymmetric transformation $\de$ realized
on both superfields $\Phi$ and $\bar\Phi$ respect the conjugacy of the
fields, in other words,
\be \overline{\de\Phi}=\de\bar\Phi.\ee
This reduces the supersymmetry algebra of the commutative model (62) to a
certain real form of the
$spl(2,1)$ algebra.
Explicitely, the $spl(2,1)$ supersymmetry transformation is given by
\be \de\Phi=
(\epsilon^+\IC_++\epsilon^-\CIC_-+\rho^{ +}\CIC_++\rho^{ -}\IC_-
+\beta Z_3 +\al^3R_3 +\al^+R_+ +\al^-R_-)\Phi\ee
and in the same way for $\bar\Phi$. The Grassmann parameters
 $\epsilon^{\pm},\rho^{\pm}$
have to fulfil
\be \overline{\epsilon^{-}}=\epsilon^+,\quad \overline{\rho^-}=
\rho^+;\ee
and the bosonic ones $\beta,\al^3,\al^\pm$
\be \bar\beta=-\beta,\quad \overline{\al^3}=-\al^3,\quad \overline{\al^+}=
-\al^-.\ee
These conditions correspond precisely to the choice of the real form of
the $spl(2,1)$ superalgebra.

 We have to check
that the constraints (65) are compatible with the $spl(2,1)$
supersymmetry.
 The most simple way to see it is to note
that

\noindent 1) the quadruples $c_\pm,z_-,c$ and $\cc_\pm,z_+,c$ form
 $spl(2,1)$ multiplets under the adjoint action in $\T$
(xy);

\noindent 2) $\{c,\Phi\}$  and $\{c,\bar\Phi\}$ trivially vanish;

\noindent 3) The constraints $\{c_\pm,\Phi\}=0$
and $\{\cc_\pm,\bar\Phi\}=0$
imply $\{Z_-,\Phi\}=0$ and $\{Z_+,\bar\Phi\}=0$, respectively.  This is
 true because of the
 explicit formulae (61).

We conclude that the model (62) on the commutative sphere is $spl(2,1)$
supersymmetric, because
  also the measure of the integral (8) is invariant with respect to the
Hamiltonian vector fields. Indeed,  one can straightforwardly check that
\be I(\{t,f\})=0\ee
for whatever $t,f\in\Ai$.

\subsection{The noncommutative case}
Here are the  ingredients needed for defining the noncommutative
deformation of the $N=2$
supersymmetric $\sigma$-model (62):

\vskip1pc
\noindent 1) The bar involution in the noncommutative case for it
coidentifies
the supersymmetry algebra and ensures the reality of the Lagrangian
$K(\bar\Phi\Phi)$.
It turns out that in the $commutative$ case the bar involution can be
expressed
in terms of  the automorphism $\circ$. Explicitely,
\be \bar c_\pm=\mp\cc_\mp,\quad \overline{\ic_\pm}=\pm\cic_\mp,\quad
\bar l_\pm=l^\circ_\mp,
\quad \bar l_3=l_3^\circ, \quad \bar c=c.\ee
Since the automorphism $\circ$  continues to make sense on the $N=2$
fuzzy supersphere, we can use the
relations (77) as the definition of the barred quantities in the
noncommutative case. Note, however,
that the barred involution $is$ $not$  an automorphism of the algebra
 $\A_n$ although its action
on the generators $l_\pm,l_3,c_\pm,\ic_\pm$ of $\A_n$ is expressible in
terms of the automorphism
$\circ$. The point  is that the bar involution acting on the product of
two generators
is not a morphism of the associative product in $\A_n$ since it is defined
by the first rule of (3), e.g.
\be \overline{c_+\ic_-}=-\overline{\ic_-}\bar c_+ =-\cic_+\cc_-.\ee
 The automorphism $\circ$, in turn, does respect the mutiplication, e.g.
\be  (c_+\ic_-)^\circ=\cc_+\cic_-.\ee
We actually use  first rule of (3) to define the barred involution of all
 elements of
$\A_n$ hence the second rule of (3)  has to be verified. For example, we
 can calculate
$\overline{(\cc_+)}$  by using the formula (48) and the first rule of
 (3). But  one  can use also  (77)
and the second rule of (3), i.e.
\be \overline{\cc_+}=\overline{\bar c_-}=-c_-.\ee
The  consistency of the definition requires that both ways of calculating
must be equivalent. Fortunately,
this is the case and we have the bar involution also
in the noncommutative case.
\vskip1pc
\noindent 2)  We need also the noncommutative analogue of the
 integral (8). It is a simple
exercise to show that the commutative measure in the variables
 $\bz,z,\bb^\al,b^\al$
can be rewritten in the variables $l_\pm,l_3,c_\pm,\ic_\pm$
as follows
\be d\bz dz d\bb^1 db^1 d\bb^2 db^2=dl_+dl_-dl_3
\de(l_+l_- +l_3^2 -\js) dc_+ dc_-d\ic_+d\ic_-.\ee
Thus we see that the measure is simply the direct product of the round
measure on the
bosonic
sphere and of the flat measure in the remaining fermionic variables.

 Upon the Berezin quantization,
the integral over the bosonic measure becomes ${1\over n+1}\Tr$
 \cite{GKP2}.
The fermionic measure, in turn,  becomes the supertrace $\ST$
(not the trace!)
over the Clifford algebra $Cf_4$. Indeed, the generators $c_\pm,\ic_\pm$
 of the Clifford
algebra satisfy the canonical anticommutation relation of a
 quantum mechanical system with two
fermionic degrees of freedom. The Clifford algebra can be identified with
 the algebra of
linear operators
acting on the corresponding Fock space. The latter is naturally graded so
 we obtain the supertrace
as \be \ST(.)\equiv \Tr(\G .),\ee
where $\G$ is the grading operator. It is a textbook fact that, upon the
quantization of the
fermionic system, the Berezin integral becomes the supertrace. It is easy
to see it directly
in the case of  one fermionic oscillator only. Then the only nonzero
Berezin integral
is the one over $c^\* c$; this is also true for the supertrace in the
 quantum case.

The integral (denote it $I_n$) in the noncommutative case has a crucial
property
\be I_n(AB-(-1)^{AB} BA)=0\ee
 for any $A,B\in\A_n$.
This property plays  the same role in the noncommutative case as  (76)
in the commutative one.
Namely it will ensure the supersymmetry of  the following action
\be S_n=I_n(K(\bar\Phi\Phi)).\ee
This is the action of the $N=2$ supersymmetric nonlinear $\sigma$-model
 on the fuzzy sphere.
The superfields $\bar\Phi$ and $\Phi$ are now elements
of the fuzzy algebra $\A_n$.
 More precisely, if we take any element
of $\A_n$, it can be written as a polynomial in the generators
$l_\pm,l_3,c_\pm,\ic_\pm$.
Now the coefficients in front of odd polynomials of the superfields have
 to anticommute with those polynomials,  for example
one has
\be \eta c_+l_+l_-=-c_+l_+l_-\eta.\ee
The coefficients in front of the even polynomials commute with them:
\be r l_-c_+c_-=l_-c_+c_- r .\ee
These rules are, of course, standard in the superworld. In the commutative
 case, $\eta$'s in (85)
were the Grassmann numbers belonging to some
Grassmann algebra $P$. In the noncommutative case, however, they are
the Grassman numbers
tensored with the grading $\G$ of the linear space $H_n$
where $\A_n$ acts.
 The tensoring with $\G$ is the representation of  $P$ (due to
$\G^2=1$)  and it ensures the correct commutative limit
of the superfields $\bar\Phi,\Phi$.
  On the other hand, $r$ in (86)
can be interpreted as a complex multiple of the unit element of $\A_n$.
An important thing  is that $\eta$'s are odd and $r$'s
even. Thus the superfields $\bar\Phi,\Phi$ are even.

The constraints in the noncommutative case are defined
by the  formulae
\be n[c_\pm,\Phi]=0,\quad n[\cc_\pm,\bar\Phi]=0.\ee
Note that the only change with respect to the commutative case
 is the replacement of the
Poisson brackets by the commutators scaled by the inverse  Planck constant
$n$. Of course,
the Hamiltonians $c_\pm,\cc_\pm$ are elements of $\A_n$.

 The $spl(2,1)$ supersymmetry transformation $\de$ is again generated by the
 $noncommutative$ Hamiltonians
$r_\pm,r_3,z_3,\ic_\pm,\cic_\pm$ given by (12),(13),(16) and (18).
 Explicitely
$$\de\Phi=n(\epsilon^+[\ic_+,\Phi]+\epsilon^-[\cic_-,\Phi]+
\rho^{ +}[\cic_+,\Phi]+\rho^{ -}[\ic_-,\Phi]$$
\be +\beta [z_3,\Phi] +\al^3[r_3,\Phi] +\al^+[r_+,\Phi] +
\al^-[r_-,\Phi]).\ee
 Here  $\al^\pm,\al^3,\beta$ are numbers
and the parameters $\epsilon^\pm$ and $\rho^\pm$ are the quantities
 of the type $\eta$  in (85).
This  is important for ensuring that a commutator of two
 supertransformations with different
coefficients is again a supertransformation.
The parameters of $\de$ are again to fulfil
the same relations (74) and (75)  as their counterparts
 in the commutative case.
With this assigment we can easily check that   we have also in
 the noncommutative case
\be \overline\de\Phi=\de\bar\Phi.\ee

One can prove that the constraints (87) are compatible
with the supersymmetry transformation
in the same way as in the commutative setting.
Summing up, we have constructed the $N=2$ $spl(2,1)$ supersymmetric
 nonlinear $\sigma$-model
on the  noncommutative sphere. Two $N=1$  $osp(2,1)$ subsupersymmetries
can be obtained
by setting respectively $\epsilon^\al=\rho^\al$ and
 $\epsilon^\alpha=-\rho^\alpha$.
\subsection{Solving the constraints}

In the flat space commutative case, one can do more than just define the
 $\sigma$-model by the
action (62) and the constraints (63). Indeed, one can effectively solve
 (63) and cast
the action (62) in terms of the solutions of the constraints. Here we are
 going to show that all this
 can be performed
also on the commutative $N=2$ supersphere and even on the noncommutative
 one. Indeed,
due to the Poisson brackets/commutation relations (34),(35)/(36),(37)
 we immediately
conclude,
that any element of $\Ai$ or
$\A_n$ of the form \be\Phi(l_\pm,l_3,c_\pm)\ee
 solves the first set of the constraints in (87) and any element
of the form \be \bar\Phi(l_\pm^\circ,l_3^\circ,\cc_\pm)\ee solves
the second set.
Let us moreover show that $every$
solution of (87) is of the form (90). Indeed, it is a simple matter to
check that every element $\Psi$
of $\Ai$
or $\A_n$ can be $unambiguosly$ written as
\be \Psi=\Phi(l_\pm,l_3,c_\pm)+
\Phi_-(l_\pm,l_3,c_\pm)\ic_-+\Phi_+(l_\pm,l_3,c_\pm)\ic_+
+\Phi_{+-}(l_\pm,l_3,c_\pm)\ic_+\ic_-.\ee
Now the fact that  $\Phi(l_\pm,l_3,c_\pm)$ is the most general solution
 of (87)
 is the direct consequence
of the Poisson brackets/commutation relations (34)-(37).
 The same argument holds also for the
 circled variables.

Finally, we remark how we can cast in components the action on the
commutative $N=2$ supersphere.
We use the fact that $l_\pm,l_3\in\Ai$ are the generators of
the ordinary sphere (cf. (32)
and (33)). We can therefore introduce variables $u, u^\*$ such that
\be l_+={ u^\*\over 1+ u^\* u},\quad l_-={u\over 1+ u^\* u},\quad l=
{ u^\* u-1\over u^\* u+ 1}.\ee
Then
\be c_+={u\bb^2+b^1\over 1+ u^\* u} \quad c_-=
{-\bb^2+ u^\* b^1\over 1+ u^\* u}.\ee
Comparing with (20), we have
\be u=z+b^1b^2,\quad  u^\* =\bz+\bb^2\bb^1\ee
and we arrive at
\be \Phi=\Phi(z+b^1b^2,\bz+\bb^2\bb^1,b^1,\bb^2).\ee
Much in the same way, we obtain
\be \bar\Phi=\bar\Phi(z+b^2b^1,\bz+\bb^1\bb^2,\bb^1,b^2).\ee
Expanding (96) and (97) in $\bb^\al,b^\al$, inserting in the action
(62) and integrating over
$b^\al,\bb^\al$, we obtain the standard
action of the $N=2$ supersymmetric $\sigma$-model in components.

\section{Outlook}
For the purpose of the quantization of the model, say by a path integral,
 it is sufficient
to work directly  in the superfield formalism.  Nevertheless, it is
perhaps of  interest to know whether one can introduce
the component fields
also in the noncommutative case. In the $N=1$ case
it turned out \cite{GKP2}
that one could not do that for a simple reason
that the $N=1$ supersphere is not the $N=0$ supersphere tensored
with some  algebra. In the $N=2$
case the  question is more subtle.  One has to  find variables "between"
the circled and the uncircled ones, like $z,\bz,b^\al,\bb^\al$
 in the commutative case,
such that the $N=2$ supersphere would be the product
 of the bosonic fuzzy sphere and the Clifford
algebra in these intermediate variables. This is  needed for
being able to take the
supertrace over the Clifford algebra separately and cast the action as
 the trace over the bosonic
fuzzy sphere only. It is an open problem; personally we feel
that it is not  possible.

Another natural question concerns coordinate transformations on the K\"ahler
target. Although for some simple manifolds
(like complex projective spaces) one can completely define
the K\"ahler potential working in  one chart,  one should anyway look for a
more invariant
definition of the theory. Of course, this problem does not concern only
 the theories on
the noncommutative worldsheets but arises in general in the studies
of  quantum
theory of the nonlinear  $\sigma$-models

\end{document}